\documentclass[aps,twocolumn,superscriptaddress,showpacs,showkeys]{revtex4}
\usepackage{graphicx}

\begin{document}

\title{Chirped chiral solitons in nonlinear Schr\"odinger equation with self-steepening and self-frequency shift}

\author{Vivek M Vyas}
\email[Email:]{vivek@iiserkol.ac.in}
\affiliation{Physical Research Laboratory, Navrangpura, Ahmedabad 380 009, India}

\author{Pankaj Patel}
\affiliation{Department of Physics, M. S. University of Baroda, Vadodara 390 002, India}

\author{Prasanta K Panigrahi}
\email[Email:]{prasanta@prl.res.in}
\affiliation{Physical Research Laboratory, Navrangpura, Ahmedabad 380 009, India}
\affiliation{Indian Institute of Science Education and Research (IISER) Kolkata, Salt Lake, Kolkata 700106, India}

\author{Choragudi Nagaraja Kumar}
\affiliation{Department of Physics, Panjab University, Chandigarh-160 014, India}

\author{W. Greiner}
\affiliation{Frankfurt Institute for Advanced Studies, Johann Wolfgang Goethe- Universit\"at, Frankfurt am Main, D-60054, Germany}

\begin{abstract}
We find exact solutions to nonlinear Schr\"odinger equation in the presence of self-steepening and self-frequency shift. These include periodic solutions and localized solutions of dark-bright type which can be {\emph{chiral}}, and chirality being controlled by sign of self steepening term. A new form of self phase modulation, which can be tuned by higher order nonlinearities as also by the initial conditions, distinct from nonlinear Schr\"odinger equation, characterizes these solutions. In certain nontrivial parameter domain solutions are found to satisfy {\emph{linear}} Schr\"odinger equation, indicating possiblity of linear superposition in this nonlinear system. Dark and bright solitons exist in both anomalous and normal dispersion regimes and a duality between dark-bright type of solution and kinematic-higher order chirping is also seen. Localized kink solutions similar to NLSE solitons, but with very different self phase modulation, are identified.
\end{abstract}


\maketitle

Nonlinear Schr\"odinger equation (NLSE):
\begin{equation}\label{nlse}
 \nonumber i {\psi}_{x} + {a_1} {\psi}_{tt} + {a_2} {|\psi|^{2}}{\psi}= 0,
\end{equation}
governs the dynamics of picosecond pulse propagation in optical fibers \cite{agrawal}, where $a_1$ is group velocity dispersion (GVD) parameter and $a_2$ specifies the strength of Kerr nonlinearity. As predicted by Hasegawa and Tappert \cite{hasegawa} and experimentally observed by Mollenauer {\em{et al.}} \cite{mollen}, this system supports stable soliton solutions owing their existence to complete integrability \cite{zs}. With the advent of high intensity laser beams, it has become possible to generate optical pulses with width of the order of $10$ femtoseconds. The higher order effects like third order dispersion, self steepening of pulse due to dependence of the slowly varying part of nonlinear polarization on time
and self frequency shift arising from delayed Raman response become important while studying propagation of these pulses. Inorder to account for them, Kodama \cite{kod} and Kodama \& Hasegawa \cite{kh} proposed higher order nonlinear Schr\"odinger equation (HNLSE) as a generalization of NLSE:
\begin{eqnarray}\label{hnlse}
\nonumber i {\psi}_{x} + {a_1} {\psi}_{tt} + {a_2} {|\psi|^{2}}{\psi}\\
+ i \left[  {a_3} {\psi}_{ttt} + {a_4} {({|\psi|^2}{\psi})}_{t}
+{a_5} {\psi}{(|\psi|^2)}_{t} \right]  = 0,
\end{eqnarray}
where a third order dispersion with coefficient $a_3$, self steepening term with coefficient $a_4$ and self frequency shift effect with coefficient $a_5$ have been added. This model, unlike NLSE, is not integrable in general. A few integrable cases have been identified: (i) Sasa-Satsuma case
($a_3$:$a_4$:$(a_4+a_5)$ = 1:6:3) \cite{ss}, (ii) Hirota case ($a_3$:$a_4$:$(a_4+a_5)$ = 1:6:0) \cite{hirota} and (iii) derivative NLSE of type I and type II \cite{ac}. Many restrictive special solutions of bright and dark type have been obtained \cite{palacios,li,cnk}.

The effect of third order dispersion is significant for femtosecond pulses when GVD is close to zero. It is negligible for optical pulses whose width is of the order of $100$ femtoseconds or more, having power of the order of 1 Watt and GVD far away from zero. However, in this case self steepening, as well as self frequency shift terms are still dominant and should be retained.
The effects of these higher order terms on pulse propagation have been extensively studied numerically \cite{agrawal,band}, and some special solutions to this system are also known \cite{olivra}.

In this letter, we report existence of a new class of localized, as well as periodic solutions for NLSE in the presence of self steepening and self frequency shift. The localized solutions include dark, bright and kink type solitons. These complex solitons are generally chirped and show a new form of self phase modulation (SPM). Unlike NLSE, where chirping is of kinematic origin (controlled by initial conditions), the chirping in present case has both kinematic as well dynamic origin. The former varies as reciprocal of intensity, whereas the latter is directly proportional to intensity and depends upon higher order nonlinearities only. It is evident that kinematic chirping plays a significant role for dark soliton, where as the dynamical chirping would be important for bright soliton.
We found that for certain special values of parameters the intensity profile of these solitons are similar to NLSE solitons, while retaining the nontrivial phase structure due to higher order nonlinear terms. It is also seen that, in the same regime, kink solitons exist which otherwise are not allowed in this system. For certain parameter values, this model mimics NLSE, however the presence of higher order nonlinearities crucially affects the dynamics. And as a consequences one sees that fundamental bright soliton of NLSE is no longer a valid solution. Further, trivial phase dark solitons exist in anomalous dispersion regime, are found to be chiral, with the direction of propagation set by self steepening term. Chirped dark and bright solitons are found to exist in normal as well as anomalous dispersion regimes. Chirped dark solitons in anomalous regime, whereas chirped bright solitons in normal dispersion regime exhibit chirality, which is controlled by self steepening term. Observation of chiral solitons is one of the main results of this letter. Very interestingly, it is seen that for some nontrivial choice of parameters, this system behaves like free particle Schr\"odinger equation with appropriate constants, making this system amenable to linear superposition, which is otherwise forbidden in this nonlinear system. This system is found to be Painlev\'{e} integrable thereby establishing existence of regular solutions.

Modulo a trivial kinematic phase, the complex envelope travelling wave solutions can be generally represented as:
\begin{equation}\label{ansatz}
\psi(x,t)=\rho(\xi)e^{i\chi(\xi)}
\end{equation}
\noindent where $\xi = \alpha (t- ux)$ is the travelling coordinate, and $\rho$ and $\chi$ are real functions of $\xi$. Here, $\alpha$ is scale parameter and $u=1/v$ with $v$ being the group velocity of the wave packet. The ansatz solution leads to the compatibility conditions:

\begin{eqnarray} \label{c1}
\nonumber - {\alpha u \rho'} + {2 {\alpha}^2 {a_1} {\chi}' {\rho}'} + {{\alpha}^2 {a_1} {\chi}'' {\rho}}  
\\+ { 3 {\alpha} {a_4} {\rho}^2 {\rho}'} + {2 {\alpha}{a_5} {\rho^2} {\rho}'}=0,
\\
{{\alpha u} {\chi}' \rho} + {{\alpha}^2 {a_1}{\rho}''} - {{{\alpha}^2 {a_1}} {{\chi}'}^2 \rho} + {a_2 {\rho}^3} - {\alpha}{a_4}{\chi}'{\rho}^3 = 0.\label{c2}
\end{eqnarray}
Equation (\ref{c1}) can be exactly integrated to yield:
\begin{equation}
\chi' = \frac{u}{2 \alpha a_1} + \frac{c}{\alpha a_1 \rho^2} - \frac{(3 a_4 + 2 a_5)}{4 \alpha a_1} {\rho}^2,\label{phase}
\end{equation}
where c is to be determined by initial conditions. It is to be noted that the phase has a nontrivial form and has two intensity dependent chirping terms, apart from kinematic first term. As is evident the second term is of kinematic origin and is common to Schr\"odinger equation as well. The last term is due to higher nonlinearities and leads to chirping that is exactly inverse to that of the former. This is a novel form of self phase modulation which is controlled by interaction. The amplitude equation (\ref{c2}) reduces to:
\begin{equation}\label{thetaeqn}
{\theta_1} \rho'' + {\theta_2} {\rho} + {\theta_3}{\rho}^3 + {\theta_4}{\rho}^5 = \frac{c^2}{\rho^3} \nonumber
\end{equation}
\noindent with $\theta_1={\alpha}^2 {a_1}^2$, $\theta_2=\frac{({u^2}-{c a_4}+{2 c a_5})}{4}$,
$\theta_3= \frac{(2 a_1 a_2 - u a_4)}{2}$ and $\theta_4=\frac{(4 a_4 - 1)(3 a_4 + 2 a_5)}{16}$.
We note that the nontrivial contribution from higher order nonlinear terms is through $\theta_4$, which is zero for $a_4=1/4$ or $a_4 : a_5 = -2 : 3$ (assuming $a_4 \neq 0$ and $a_5 \neq 0$), and as we shall soon show, this results in interesting physical consequences. 
In the case when $a_4 : a_5 = -2 : 3$, both the intensity as well as phase will not be having any new features due to higher order nonlinearities and the solutions will exactly resemble NLSE solutions. However unlike NLSE both dark and bright solitons exists, in both normal and anomalous dispersion regime. When $c=0$, existence of fundamental bright solitons with $\rho = A \mathrm{sech(\xi)}$ is forbidden, and only dark solitons with $\rho= A \mathrm{tanh(\xi)}$ exist, with ${\alpha}^{2}=\frac{u^2}{2 {a_1}^2}$. Furthermore, these dark solitons in anomalous dispersion regime, i.e., $a_1 > 0$, respect the inequality ${a_4} u \geq 2 |{a_1}| |{a_2}|$,  which restrict them to travel only along one direction, given by sign of $a_4$. 
This is an example of chiral soliton which is absent in NLSE.
When $c \neq 0$, both bright soliton and dark soliton exists, and satisfy $2 a_1 a_2 > u a_4$ and $2 a_1 a_2 < u a_4$ respectively, showing that both have mutally exclusive velocity space. As a consequence, in anomalous dispersion regime, dark solitons obey the inequality $2 |a_1| |a_2|< u {a_4}$ hence are chiral, whereas in normal dispersion regime bright solitons satisfy $-2 |a_1| |a_2| > u {a_4}$ and hence are also chiral. Notice that the directionality of these solitons is due to the presence of higher order terms, the sign of $a_4$ decides the direction in which solitons are allowed to propagate. For $a_4=1/4$, the intensity profile will be the same as NLSE, whereas phase will still show nontrivial chirping. In this case also both dark and bright solitons can be chiral, and can exist in normal and anomalous disperison regimes, which is in sharp contrast to NLSE.

It is very intriguing to see that when $a_{4}=\frac{-2 a_{5}}{3}$ and $u=\frac{-3 a_{1} a_{2}}{a_{5}}$, equation (\ref{thetaeqn}) combined with equation (\ref{phase}) reduces to free particle Schr\"odinger equation in $\psi$. So for this choice of parameters, in the presence of both Kerr and higher order nonlinearities, the effective evolution equation for $\psi$ is linear, and one would expect to see phenomena like interference, which is forbidden otherwise in this system.

Equation (\ref{thetaeqn}) can be cast into a convenient form using $\rho=\sqrt{\sigma}$:
\begin{equation}
{\frac{\theta_1}{2} \sigma''} + 2 \theta_2 \sigma + {\frac{3 \theta_3}{2}}{\sigma^2} + {\frac{4 \theta_4}{3}}{\sigma}^3 = k,\label{inten}
\end{equation}
where k is constant fixed by initial conditions. Solutions for this equation, with $\theta_4 \neq 0$, can be found by conformal M\"obius transformation:
\begin{equation}
\sigma = \frac{A + B f}{C + D f},\label{solun}
\end{equation}
which for some suitable $A$, $B$, $C$ and $D$, connects $\sigma$ to elliptic function $f$. These elliptic functions, as is known, are generalization of trigonometric and hyperbolic functions and appear in solutions of many nonlinear equations.

Considering the importance of localized solutions, we set $f(\xi)=sech(\xi)$, and look for allowed values of $A$, $B$, $C$ and $D$ for which (\ref{solun}) is solution of equation (\ref{inten}). The consistency conditions leads to: $A=8\theta_4 \tilde{A} -3 \theta_3$,
$B=8\theta_4\tilde{B}-3\theta_3\tilde{D}$, $C=8\theta_4$ and $D=8\theta_4 \tilde{D}$
where $\tilde{A}$, $\tilde{B}$, $\tilde{D}$ and $\alpha$ are given by:
$(1024 {\theta_4}^4) {\tilde{A}}^3 + (1536 \theta_2 {\theta_4}^3 + 432 {\theta_3}^3 {\theta_4}^2 - 864 {\theta_3}^2 {\theta_4}^2) {\tilde{A}} + (-3 {\theta_3}^4 + 162 {\theta_3}^3 \theta_4 - 576 \theta_2 \theta_3 {\theta_4}^2 - 768 k {\theta_4}^2)=0$, $\tilde{B}=\frac{-\tilde{D} (-54 {\theta_3}^2+27 {\theta_3}^3 + 96 \theta_2 \theta_4 +128 {\theta_4}^4 {\tilde{A}}^2)}{64 \tilde{A}{\theta_4}^2}$, $\tilde{D}=\pm \frac{8 \sqrt{2} \tilde{A} \theta_4}{\sqrt{54 {\theta_3}^2 - 27 {\theta_3}^3 - 96 \theta_2 \theta_4 - 64 {\theta_4}^2 {\tilde{A}}^2 }}$ and  
${\alpha}^2 = \frac{18 {\theta_3}^2 - 9 {\theta_3}^2 - 32 \theta_2 \theta_4 - 64 {\tilde{A}} {\theta_4}^2}{8 {a_1}^2 \theta_4}$.


\begin{figure}
\includegraphics[scale=0.5]{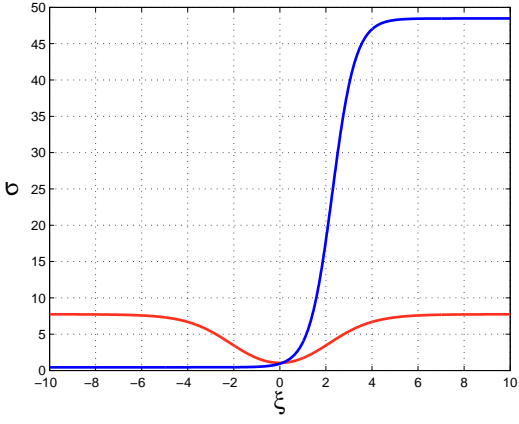}\caption{\label{intensoln}Intensity profile of few solutions: i) Dark Soliton (in red color) for $a_1=1.6001$, $a_2=-2.6885$, $a_3=-2.8302$, $a_4=0.30814$, $a_5=0.76604$, $u=4.1185$, $c=-3.1186$, $k=-77.965$, $A=35.36$, $\alpha=4.7421$, $B=-11.912$, $C=4.5702$ and $D=17.855$; ii) Kink Soliton (in blue color)
$a_1=-11.197$, $a_2=44.778$, $a_3=6.219$, $a_4=19.066$, $a_5=37.301$, $u=884.36$, $c=-13810$,
$k=4360.2$, $A=0.94014$, $\alpha=8.6548$, $B=0.06745$ and $D=-0.97921$.}
\end{figure}
%

\begin{figure}
\includegraphics[scale=0.5]{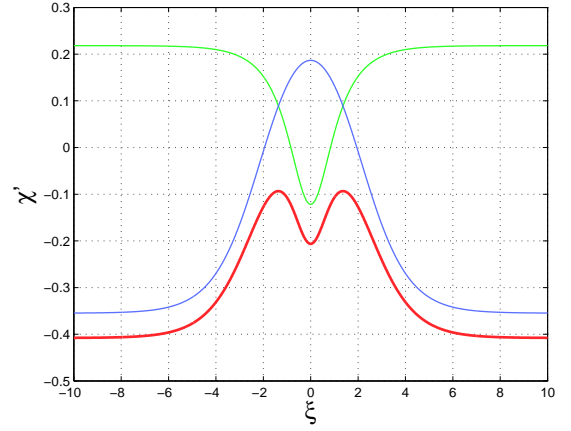}\caption{\label{sechphase}Phase profile of dark soliton plotted in Figure 1. (in red color). Blue curve shows the contribution from linear chirp whereas green curve shows the contribution from higher order chirp. The chirp reversal is clearly seen as peaks in the red curve.}
\end{figure}

Since the exact closed form solution is known, a simple maxima-minima analysis is sufficient to distinguish parameter regimes supporting dark/bright solitary waves \cite{pkp2}. In this case, when $AD>BC$ one gets a bright soliton, whereas if $AD<BC$ then dark soliton exists.
Figure (\ref{intensoln}) depicts intensity profile of a typical dark soliton.
It is interesting to note that for dark solitons, from equation (\ref{phase}), the kinematic chirping is dominant at the centre of the pulse whereas the higher order chirping is
dominant away from the center (see figure (\ref{sechphase})). However, exactly the opposite is true for bright solitons where the center is dominated by higher order chirping and kinematic chirping is important away from the center. This shows that there is duality between dark soliton-kinematic chirping and bright soliton-higher order chirping.
A mutual cancellation will occur at some point(s) when both kinematic chirping and higher order chirping are comparable and have opposite signs, and will result in chirp reversal at the point(s) of cancellation.
Chirp reversal plays a significant role in fiber optics, and has attracted considerable attention recently in context of pulse retrieval in dispersion-nonlinearity managed optical fibres \cite{moores,kruglov,kumar,atre}.
Chirp reversal occurs at $\xi_r = \pm \mathrm{cosh}^{-1} (\frac{D \sigma_c - B}{A - \sigma_c C})$ provided $\frac{-4 c}{3 a_4 + 2 a_5}>0$ and $\frac{D \sigma_c - B}{A - \sigma_c C} \geq 1$, where $\sigma_c=\sqrt{\frac{-4 c}{3 a_4 + 2 a_5}}$. We have plotted $\chi'$ against $\xi$ in figure (\ref{sechphase}) where the chirp reversal is clearly seen as two maxima in the profile.

It should be noted that equation (\ref{inten}) with $\theta_4\neq 0$ has no kink solutions, which are of the type:
\begin{equation}
 \sigma = \frac{A + B \mathrm{tanh(\xi)}}{C + D \mathrm{tanh(\xi)}}.
\end{equation}
However, for $a_4=1/4$, $\theta_4$ is zero allowing existence of these kind of solutions.
In this case, $k=-\frac{({\theta_1}^2 + 2 {\theta_2}^2)}{3 \theta_3}$,
$A=-\frac{-2 ({\theta_1} + {\theta_2}) \pm \sqrt{12 ({\theta_1}^2 - {\theta_2}^2) - 18 \theta_2 k}}{3 \theta_3}$, $B=\Gamma D$ where $\Gamma=-\frac{(2 \theta_1 A + 3 \theta_3 A^2 + 8 \theta_2 A - 6k)}
{6 \theta_3 A + 4 \theta_2 - 2 \theta_1}$, $C=1$ and $D=\pm \sqrt{ \frac{2k - 4 \theta_2 A - 3 \theta_3 A^2}{2 \theta_1 (A - \Gamma)} }$.
Figure (\ref{intensoln}) shows intensity profile of a typical kink solution.
These solutions being asymmetric around $\xi=0$, have interesting phase profile which show chirp reversal only once along the profile. The condition for existence of this reversal is given by: $\frac{-4 c}{3 a_4 + 2 a_5}>0$ and $-1 \leq \frac{A - \sigma_c C}{D \sigma_c - B} \leq 1$, and the point of reversal is $\xi_{r} = \mathrm{tanh^{-1}}(\frac{A - \sigma_c C}{D \sigma_c - B})$.


Solutions to equation (\ref{inten}) via (\ref{solun}) are not restricted to localized ones alone, periodic solutions also exist. Infact it is easy to show that:
\begin{equation}
\sigma = \frac{-3 \theta_3}{8 \theta_4} \pm \frac{C \mathrm{cos(\xi)}}{\sqrt{2} \pm \mathrm{cos(\xi)}}
\end{equation}
is a periodic solution of equation (\ref{inten}), provided ${\theta_1}^3 +
2 {\theta_1}^2 (2 \theta_2 - \frac{9 {\theta_3}^2}{8 \theta_4} + \frac{9 {\theta_3}^2 }{16 \theta_4}) + \frac{8 \theta_4}{3} {(\frac{-3 \theta_2 \theta_3}{2 \theta_4} + \frac{27 {\theta_3}^3}{64 {\theta}^2} - \frac{{\theta_3}^4}{128 {\theta_4}^3})}^{2} = 0$ and $C=-2 {\theta_1}^2 [2 \theta_2 - \frac{9 {\theta_3}^2}{8 \theta_4} + \frac{9 {\theta_3}^2 }{16 \theta_4}]$.

Apart from the solutions discussed above, amplitude equation (\ref{thetaeqn}) albeit with different parameters, has been carefully studied in context of cubic quintic nonlinear Schr\"odinger equation \cite{winter1,schurman}. It has been shown that this equation possesses a rich solution space, where the solutions are expressible in terms of Weierstrass functions, and nature of the solution crucially depends upon initial conditions. Similar analysis for this system would be relevant, and will shed light on the structure of solution space.

A natural question arises whether the model is integrable in this regime or not. Following the Ablowitz-Ramani-Segur algorithm, we investigate singularity structure of the ordinary differential equation (\ref{inten}), which is obtained from an exact reduction of the original partial differential equation (\ref{hnlse}) \cite{ablo,gr}. Interestingly, we found that ordinary differential equation represented by (\ref{inten}) possess poles as the only movable singularities, which implies that this system indeed has the P-property \cite{ablo,gr}. Hence, we see that this system passes Painlev\'{e} test, and is Painlev\'{e} integrable, which guarantees existence of regular solutions in general.

In conclusion, we have found a new class of exact solutions to NLSE system in the presence of self steepening and self frequency shift terms. These include localized solutions of dark-bright type, kink solutions and periodic solutions. These solutions have nontrivial phase chirping which varies as a function of intensity and are different from that in Ref. \cite{palacios} where the solutions had a trivial phase. A nontrivial connection of this sytem with linear Schr\"odinger equation in appropriate limits is pointed out. A duality is seen between the dark-bright type of solution and kinematic-higher order chirping. A novel form of self phase modulation has been observed in this case, which shows chirp reversal across the pulse profile. It is known that prechirping of pulses often leads to better quality of pulses, in particular it is quite effective with distributed GVD and nonlinearity \cite{moores,turit1,turit2,atre}. In this context, the solutions having chirping due to initial conditions as well as dynamical conditions will provide a better control.
It is noted that for some parameter values the intensity and phase of these solitons will exactly be the same as NLSE solitons, and are found to be chiral, with the direction of propagation controlled by self steepening term. Both dark-bright solitons are found to exist in normal-anomalous dispersion regimes. It is seen in some cases that intensity of these solitons would be like NLSE soliton and only phase structure will be different. Kink solutions are found to exist in this system for special choice of parameters. The system is seen to possess P-property and hence is Painlev\'{e} integrable.

\begin{acknowledgments}
CNK would like to thank DAAD for the research stay in Germany and FIAS for warm hospitality. Part of the work was also supported by CSIR, India, through a research project.
\end{acknowledgments}

\end{document}